Thin Film Substrates from the Raman spectroscopy point of view

L. Gasparov[1], T. Jegorel[1,2], L. Loetgering[1,3], S. Middey[4], J. Chakhalian[4]

[1]Department of Physics, University of North Florida, Jacksonville, Florida, 32224, USA

[2]University of Technology, 10010, Troyes, France

[3]Fraunhofer Institute for Laser Technology, 52074, Aachen, Germany

[4]Department of Physics, University of Arkansas, Fayetteville, Arkansas, 72701, USA

We have investigated ten standard single crystal substrates of complex oxides on the account of their applicability in the Raman spectroscopy based thin film research. In this study we suggest a spectra normalization procedure that utilises a comparison of the substrate's Raman spectra to those of well-established Raman reference materials. We demonstrate that MgO, $LaGaO_3$, $(LaAlO_3)_{0.3}(Sr_2AlTaO_6)_{0.7}$ (LSAT), $DyScO_3$, $YAlO_3$, and $LaAlO_3$ can be of potential use for a Raman based thin film research. At the same time $TiO_2$ (rutile), $NdGaO_3$, $SrLaAlO_4$, and $SrTiO_3$ single crystals exhibit multiple phonon modes accompanied by strong Raman background that substantially hinder the Raman based thin film experiments.

## I. INTRODUCTION

Thin film research is an integral part of modern materials science. Investigation of the artificial heterostructures consisting of correlated oxide materials represents one branch of such studies. This research gives insight into complex interactions taking place in the correlated oxides and phenomena associated with the stress present on the interface of the oxide and the substrate. Artificial heterostructures have a potential to yield a number of electronic applications. [1-3].

Optical spectroscopy in general and Raman spectroscopy in particular provide a versatile non-destructive tool to investigate thin films. For instance, Raman spectroscopy allows one to probe low frequency elementary excitations such as phonons, magnons and some electronic excitations as well as complex interplay between them. Many ccorrelated phenomena such as metal-insulator transition, orbital ordering, and charge ordering are manifested through changes in structural symmetry[4-7]. These changes can be probed by Raman technique. Furthermore Raman spectroscopy has become one of the leading tools to investigate oxide heterostructures as the substrate induced strain effect causes additional structural changes.[8-10] All this makes Raman spectroscopy a valuable tool to investigate correlated phenomena in thin films.

A choice of a thin film substrate is dictated by minimisation of the mismatch between the film and the substrate’s unit cell. However such choice often overlooks Raman properties of the substrate and may unintentionally render such substrate useless for a Raman-based experiment particularly when the substrate exhibits rich and intensive Raman spectrum. Establishing a procedure for selection of a good Raman substrate is one of the goals of this study.

In a Raman experiment the laser light is shone on the film and scattered light is analysed. Typically the film is too thin to prevent the light from penetrating into the substrate. As a consequence the Raman spectra of both the film and the substrate are detected in the experiment. In majority of the situations only the spectrum of the film is of interest to a researcher. Hence it is necessary to separate the spectrum of the film from that of the substrate.

A vast majority of the film's low energy excitations is typically located below eight to nine hundred wavenumbers (<100 meV) Raman shift. A substrate that has no Raman modes within this frequency range and exhibits low Raman background will not mask any excitations that come from the film. Such substrate would be considered a good candidate for a Raman based thin film experiment. The goal of this study is to select such substrate. This goal can be achieved through comparison of the intensities and the frequencies of the substrate's modes.

Comparing the intensity of Raman signal of different materials is often a challenge. The use of an absolute scattering cross-section in Raman spectroscopy is rare. In most of the situations the researchers are interested in the Raman shift of a certain mode but not in its absolute cross-section.[11] As a consequence the Raman intensity is routinely reported in arbitrary units. Different sensitivity of the detectors, different collection optics, and different configurations of experiment performed by different research groups make it practically impossible to make direct comparison of the raw spectra reported in the literature. However, evaluation of the normalised Raman intensity provides a sensible alternative to the absolute scattering cross-section measurements. One of the oldest methods that have been used is the sample substitution method[11,12]: during the experiment the sample is replaced by a standard by "means of a lateral displacement"[11,12] and spectra of both the sample and the reference are recorded and compared. The wide gap insulators with charge gap exceeding the energy of the visible light photon are of particular use as the references. In particular, $CaF_2$ ($E_g$=12 eV)[13], $BaF_2$ ($E_g$=11eV)[13], and diamond ($E_g$=5.5eV)[14]can be used for this purpose.[12]

In this short paper we suggest a modified sample replacement procedure that allows one to make a consistent comparison of the Raman intensities of different materials. We employed this procedure to analyse different single crystal complex oxide substrates and found that $LaAlO_3$, $YAlO_3$, $LaGaO_3$, $DyScO_3$, $(LaAlO_3)_{0.3}$ $(Sr_2AlTaO_6)_{0.7}$ (LSAT), MgO substrates can be used in a Raman spectroscopy based thin film study while $NdGaO_3$, $SrLaAlO_4$, $SrTiO_3$, and $TiO_2$ (rutile) display strong background signal and/or too many phonon modes to be of practical use in such Raman experiment.

## II. EXPERIMENTAL

Un-polarized Raman spectra of commercially available substrates (Crystec, Germany) have been measured at room temperature using 514 nm line of Coherent Innova-70 Argon Ion Laser with the laser power not exceeding 10 mW when focused on the sample. The spectra were collected in the backscattering geometry using microscope attachment of the Horiba/Jobin Yvon T64000 Raman spectrometer operated in the double subtraction mode. All the measurements were performed with Olympus × 100 microscope objectives with numerical aperture of 0.9. The objective produced a 2 µm laser spot on the sample. Corresponding laser intensity did not exceed 2.5 $GW/m^2$ or 250 $kW/cm^2$.

## III. RESULTS AND DISCUSSION

In the experiment we compare the Raman intensity of a sample to that of $CaF_2$ and Si. $CaF_2$ displays a well pronounced Raman mode at 321 $cm^{-1}$. The $CaF_2$ charge gap[13] of 12 eV is significantly larger than the visible photon's energy. This fact assures that no resonance effects will affect the Raman scattering in the $CaF_2$. Si is a very common reference material in practically every Raman laboratory. This material displays a well pronounced single mode at 520 $cm^{-1}$ which is routinely used for a Raman setup alignment. This is the reason to provide the spectra normalised to the Si in addition to those normalised to $CaF_2$ even though Si has a charge gap in the infrared range (1.1 eV)[15] and the resonance effects in Si may not be neglected.

Normalization of the Raman intensity of a material of interest to that of known Raman reference allows one to make a consistent comparison of Raman intensities irrespective of the experimental setup, the type of detector, and the configuration of the experiment. Figs. 1 through 3 illustrate suggested approach. In particular, Fig.1 displays how the Raman spectra of Si (Fig.1 a) and $CaF_2$ (Fig.1 b) change with the laser power. The insets in the figure display the peak intensity (520 $cm^{-1}$ mode in Si, and 321$cm^{-1}$ mode in $CaF_2$) vs. laser intensity. The slope of such graph provides one with the intensity of a particular spectral feature expressed in CCD counts per second per $W/m^2$ of laser intensity. The measurements of Silicon and $CaF_2$ yielded the slopes of $m_{\mathrm{Si}}=(45.5\pm1.1)*10^{-8}\ \frac{CCD\ counts}{\frac{W}{m^2}s}$and $m_{CaF2}= (2.8\pm0.1)*10^{-8}\ \frac{CCD\ counts}{\frac{W}{m^2}s}$, respectively. Throughout the paper we will call these values of these slopes the m-values. We utilise these m-values as the references for our comparison procedure.

We would like to note that the use of the slopes of the peak intensity vs. incident laser intensity eliminates a potential effect of a non-Raman background. In other words, if there is an offset in the Raman spectrum due to a flat non-Raman background then such offset will not affect the slope of this graph but will show up as a non-zero intercept on the peak intensity vs. laser intensity graph.

Fig. 2 illustrates how a spectrum of a substrate can be analysed. The top panel of Fig.2 displays how the Raman spectrum of $DyScO_3$ substrate changes with laser power. There are many phonon modes in the spectrum with the most prominent mode situated at 155$cm^{-1}$. The inset displays linear increase of the mode's peak intensity with the incident laser power yielding corresponding m-value of $m$= (3.31±0.06)*10$^{-8}$ $\frac{CCD\ counts}{\frac{W}{m^2}s}$.

The lower panel of Fig.2 displays the normalised intensity of the substrate's spectrum. It is calculated by first dividing the spectrum intensity to that of the most dominant spectral feature. In $DyScO_3$ this is the 155-cm$^{-1}$ mode. The resultant is then multiplied by the ratio of the $m$-value of the most prominent mode of the substrate to that of the reference. The ensuing equation is as follows: $I_{norm} = \frac{I(\omega)}{I_{max}} * \frac{m_{max}}{m_{ref}}$where I(ω) is the Raman spectrum of the material of interest as measured by the detector, $I_{max}$ is the peak intensity of the most prominent spectral feature of the material of interest, $m_{max}$ is the slope of the graph of the intensity of the most prominent spectral feature in the material of interest vs. incident laser intensity, $m_{ref}$ is that of the reference material. In principle one could just divide the Raman intensity of the substrate by the intensity of the main feature in the reference's spectrum. This however would lead to much larger uncertainty compared to the use of slopes.

The lower panel of Fig.2 displays the result of such normalization for the spectra measured with different laser power. These spectra nearly perfectly overlap with each other which is in our opinion substantiates the validity of the approach. One does observe a slight deviation from perfect overlap at the high wavenumber end of the spectra beyond 900 cm$^{-1}$. Such deviation arises due to relatively larger experimental error for the spectra measured with the low laser power.

Another important feature of a good Raman substrate's spectrum is a spectral range that is free from any strong Raman modes. The extent of this Free Spectral Range (FSR) is one of the key characteristics of a useful Raman substrate. In $DyScO_3$ the FSR extends from 570 to 900 wavenumbers, Fig.2. The normalised (to Si) intensity of the $DyScO_3$ Raman background in the FSR is about 0.1. This value is significantly smaller than that in Si. FSR broader than 200 wavenumbers would be critical for any substrate to be used in the Raman base thin film measurements. We consider materials with broad FSR and weak Raman background in FSR to be potentially useful Raman substrates.

Table I summarises the results of the analysis. The first three columns of the table list the substrates that were measured, the Raman frequencies of the main spectral lines that are detected within the first 900 wavenumber range, and the Raman frequency of the dominant mode in the spectrum. The m-number column lists corresponding m-values and Figs. 1 and 2 illustrate the procedure that yields these numbers. Two following columns display the normalised intensities of the dominant mode: the $I/I_{Si}$ column lists the intensity normalised to that of the 520 $cm^{-1}$ mode in Si, and the $I/I_{CaF2}$ column display the intensity normalised to the 320 $cm^{-1}$ $CaF_2$ mode. The free spectral range (FSR) column displays the frequency range that is free of Raman modes. Two following columns display the normalised (to both Si and $CaF_2$ modes) intensity of the Raman background within the FSR. The very last column displays the values of band gap in the investigated material.

Our analysis does not take into account corrections for the index of refraction of the substrate, corrections arising from the differences in the frequencies of the reference modes as well as those coming from the differences in the Raman scattering length. Such corrections are indeed necessary for obtaining the Raman scattering cross-sections. However the purpose of this report is not to obtain such cross-section values but rather gage usefulness of certain substrates for a Raman based thin film experiment.

The goal of the analysis is to find which substrate may be of potential use for a thin film research involving Raman technique. We define a good Raman substrate as one that has an FSR broader than 200 $cm^{-1}$ and with the normalised (to $CaF_2$) intensity in FSR not higher than 0.65. The later number is the FSR Raman intensity observed in Si.

Based on these criteria we found magnesium oxide (MgO) to be the best Raman substrate among those that we have measured. This material does not display any strong Raman modes between 200 and 900 wavenumbers and the relative (to $CaF_2$) background intensity in the FSR is about 0.24 as opposed to 0.65 in Si or 2.9 in strontium titanate (Table I).

Aside of MgO all other substrates that we measured display a number of phonon modes and/or broad Raman background. Some of these substrates display FSR broader than 200$cm^{-1}$ with low Raman background intensity. Those are $LaGaO_3$, LSAT, $DyScO_3$, $YAlO_3$, and $LaAlO_3$. These substrates may have a potential use for the Raman based thin film research. However actual use of any of these substrates depends on the particular frequency of the spectral features of interest in the spectrum of the film. If these features fall within the FSR of the substrate then the substrate will be of use. Table I may be of help when deciding on a particular substrate.

In the situation when reasonable FSR is present the Raman intensity within the FSR seems to correlate with the extent of the substrate's charge gap ($E_g$). Table I lists the values of the charge gap for all the measured materials and Fig. 3 displays how the relative Raman intensity within the FSR varies with the value of $E_g$. When the photon energy of the exciting laser radiation exceeds the value of charge gap a significant Raman background appears in the spectrum. The substrates with the charge gap larger than the photon energy of the exciting laser radiation display nearly identical Raman background. A broad band gap is one of the characteristic of a good Raman substrate.

When number of phonon modes in the spectrum is large the FSR becomes small (<200 $cm^{-1}$). In such a situation it is difficult to separate phonon modes from the Raman background and the substrates would be a poor choice for a Raman based film study. The substrates that display insufficiently small FSR include $TiO_2$ rutile, $NdGaO_3$, $SrLaAlO_4$, and $SrTiO_3$. The letter material is by far the worst Raman substrate with practically no FSR and high background intensity over 100-900 wavenumbers spectral range.

## IV. CONCLUSION

To summarize, we suggest a method of comparison of the Raman intensity of different materials that utilises known Raman references such as Si and $CaF_2$. We employed this method to compare a number of thin film substrates on the account of their usefulness for a Raman based thin film research. In this study we measured MgO, $LaGaO_3$, LSAT, $DyScO_3$, $YAlO_3$, $LaAlO_3$, $TiO_2$ rutile, $NdGaO_3$, $SrLaAlO_4$, and $SrTiO_3$.

Based on our analysis MgO is the best Raman substrate with largest free spectra range and fairly low Raman background. $LaGaO_3$, LSAT, $DyScO_3$, $YAlO_3$, $LaAlO_3$ can be considered as good Raman substrates since they display a reasonably wide spectral range that is free of the phonon modes together with low intensity flat Raman background within this range.

Finally $TiO_2$ rutile, $NdGaO_3$, $SrLaAlO_4$, and $SrTiO_3$ display a very little of free spectral range and thus the interpretation of Raman spectra with those substrates is rather complex and may dramatically hinder the physics associated with the films of interest. The strontium titanate (STO) is by far exhibits the strongest Raman response among those that we measured. Specifically, it has practically no FSR and displays high background intensity over 100-900 wavenumbers.

ACKNOWLEDGEMENTS

L.G. acknowledges the support from the National Science Foundation (NSF) Grants DMR-0805073, DMR-0958349, Office of Naval Research award N00014-06-1-0133 and the UNF Terry Presidential Professorship. J. C. was supported by DOD-ARO under Grant No. 0402-17291.

Table I. Raman characteristics of the references and substrates

| Substrate | Dominant spectral features ($cm^{-1}$) | The main mode<br>$cm^{-1}$ | The main mode's m-number<br>$10^{-8}\frac{\mathbf{cnts}}{\frac{\mathbf{W}}{\mathbf{m^2}}\cdot \mathbf{s}}$ | $I/I_{Si}$ | $I/I_{CaF2}$ | Free spectral range (FSR)<br>$cm^{-1}$ | FSR back-ground intensity<br>$I_{FSR}/I_{Si}$ | FSR back-ground intensity<br>$I_{FSR}/I_{CaF2}$ | $E_g$<br>eV |
|---|---|---|---|---|---|---|---|---|---|
| Si | 520 | 520 | 45.5±1.1 | 1.0 | 16 | 100-280<br>320-450<br>550-900 | 0.04 | 0.65 | 1.11[15] |
| $CaF_2$ | 321 | 321 | 2.8±0.1 | 0.06 | 1.00 | 200-300<br>350-900 | 0.017 | 0.27 | 12[13] |
| $TiO_2$ rutile | 234, 448, 612 | 612 | 59.9±2.7 | 1.31 | 21.22 | 750-900 | 0.05 | 0.8 | 3[16] |
| $LaAlO_3$ | 32, 123 | 32 | 36.9±4.5 | 0.81 | 13.1 | 180-475<br>510-900 | 0.01 | 0.19 | 5.6[17] |
| $SrTiO_3$ | 248, 305, 353, 621, 680, 716 | 305 | 21.2±0.4 | 0.46 | 7.5 | 100-170<br>500-590 | 0.18<br>0.08 | 2.9<br>1.3 | 3.2[18] |
| $LaGaO_3$ | 59, 104, 120, 150 | 120 | 11.9±0.3 | 0.26 | 4.23 | 480-900 | 0.01 | 0.16 | 4.4[19] |
| LSAT | 467, 877 | 877 | 6.3±0.2 | 0.14 | 2.3 | 170-400<br>660-830 | 0.013 | 0.21 | >2.4 |
| $NdGaO_3$ | 213, 448, 520, 732 | 213 | 5.1±0.2 | 0.11 | 1.8 | 780-900 | 0.022 | 0.36 | 3.8[20] |
| $DyScO_3$ | 157, 308, 326, 355, 458, 474, 508 | 157 | 3.33±0.06 | 0.07 | 1.2 | 570-900 | 0.011 | 0.18 | 5.9[21] |
| $YAlO_3$ | 148, 283, 343 | 282 | 3.27±0.12 | 0.07 | 1.1 | 600-900 | 0.01 | 0.16 | 8.8[22] |
| $SrLaAlO_4$ | 323, 443, 665 | 443 | 1.13±0.12 | 0.024 | 0.40 | 780-900 | 0.018 | 0.29 | 2.8[23] |
| MgO | No Raman modes, flat background | | 0.69±0.06 | 0.015 | 0.24 | 100-900 | 0.015 | 0.24 | 7.8[24] |

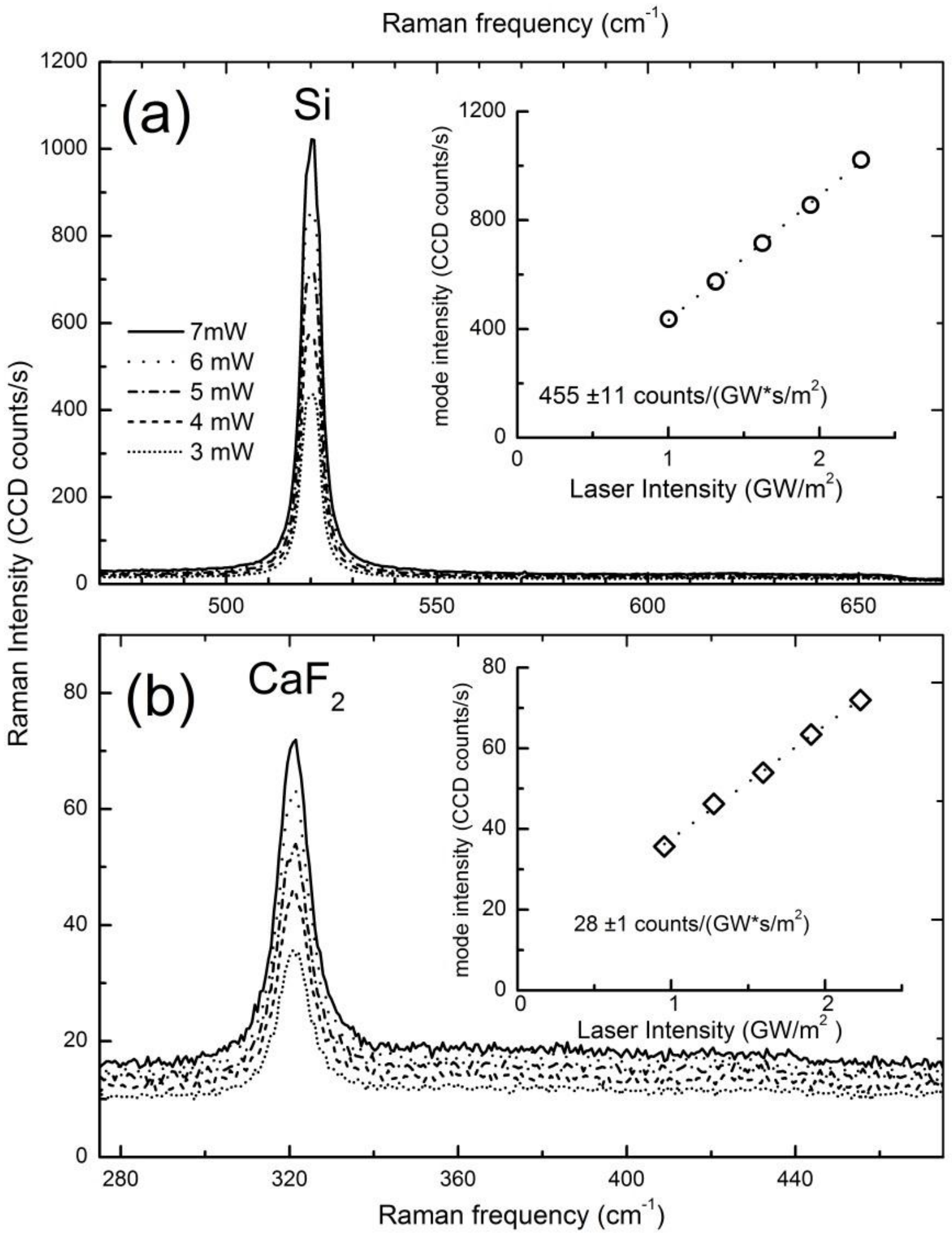

Fig.1 Raman spectra of Si (a) and $CaF_2$ (b) as a function of incident laser power. The laser was focused in 2 micron spot and the laser power on the sample varied between 3 and 7 mW. Fig.1(a) inset displays how the Raman intensity of the 520 $cm^{-1}$-Si mode increases with the intensity of incident laser radiation. The slope of the graph yields $m_{\mathrm{Si}}$=(455±11) $\frac{CCD\ counts}{\frac{W}{m^2}s}$

Fig.1(b) inset displays how the Raman intensity of the 320 $cm^{-1}$-$CaF_2$ mode increases with the intensity of incident laser radiation. The slope of the graph yields $m_{CaF2}$= (28±1) $\frac{CCD\ counts}{\frac{W}{m^2}s}$

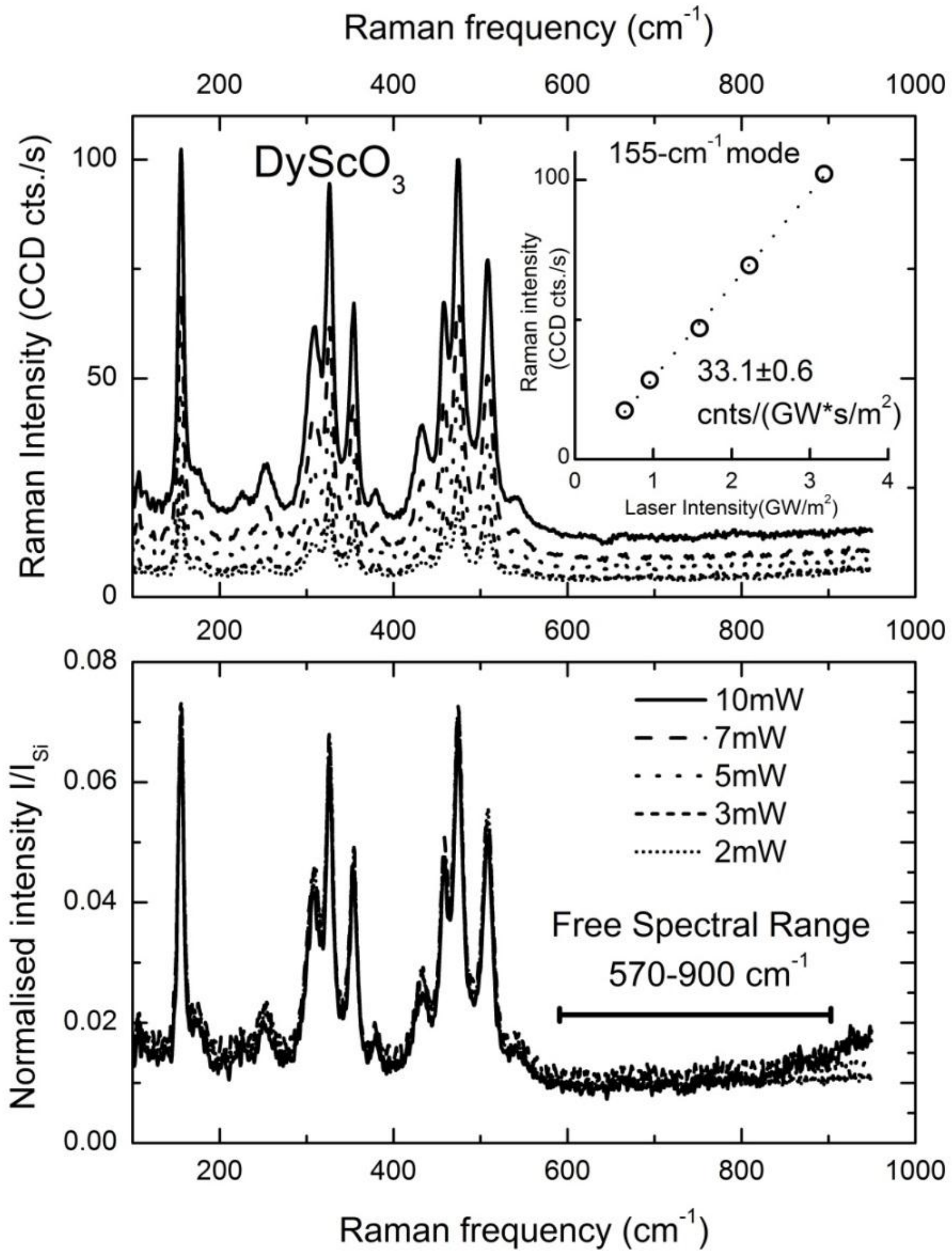

Fig.2 Normalization procedure for the Raman spectra of $DyScO_3$: The laser was focused in 2 micron spot and the laser power on the sample varied between 3 and 10 mW. The top panel displays the Raman spectra of $DyScO_3$ as a function of incident laser power. The dominant mode is located at 155 $cm^{-1}$. Top panel inset displays how the intensity of dominant Raman mode increases with the intensity of incident laser radiation. The linear dependence yields the slope $m_{155}= 33.1 \pm 0.6$ CCD counts/(GW*s/$m^2$).

The lower panel displays the spectra of $DyScO_3$ normalized to the intensity of the Si 520 $cm^{-1}$ mode according to the $I_{norm} = \frac{I(\omega)}{I_{max}} * \frac{m_{155}}{m_{Si}}$ where I(ω) is the $DyScO_3$Raman spectrum intensity as measured by the CCD detector, $I_{max}$ is the peak intensity of the 155 $cm^{-1}$ mode, $m_{155}$ is the slope of the graph in the inset of this figure, $m_{Si}$ is that of the Si, Fig.1(a) inset. Note that the spectral range between 570 and 900 wavenumber is free of any strong Raman modes with the normalised Raman background intensity around 0.1.

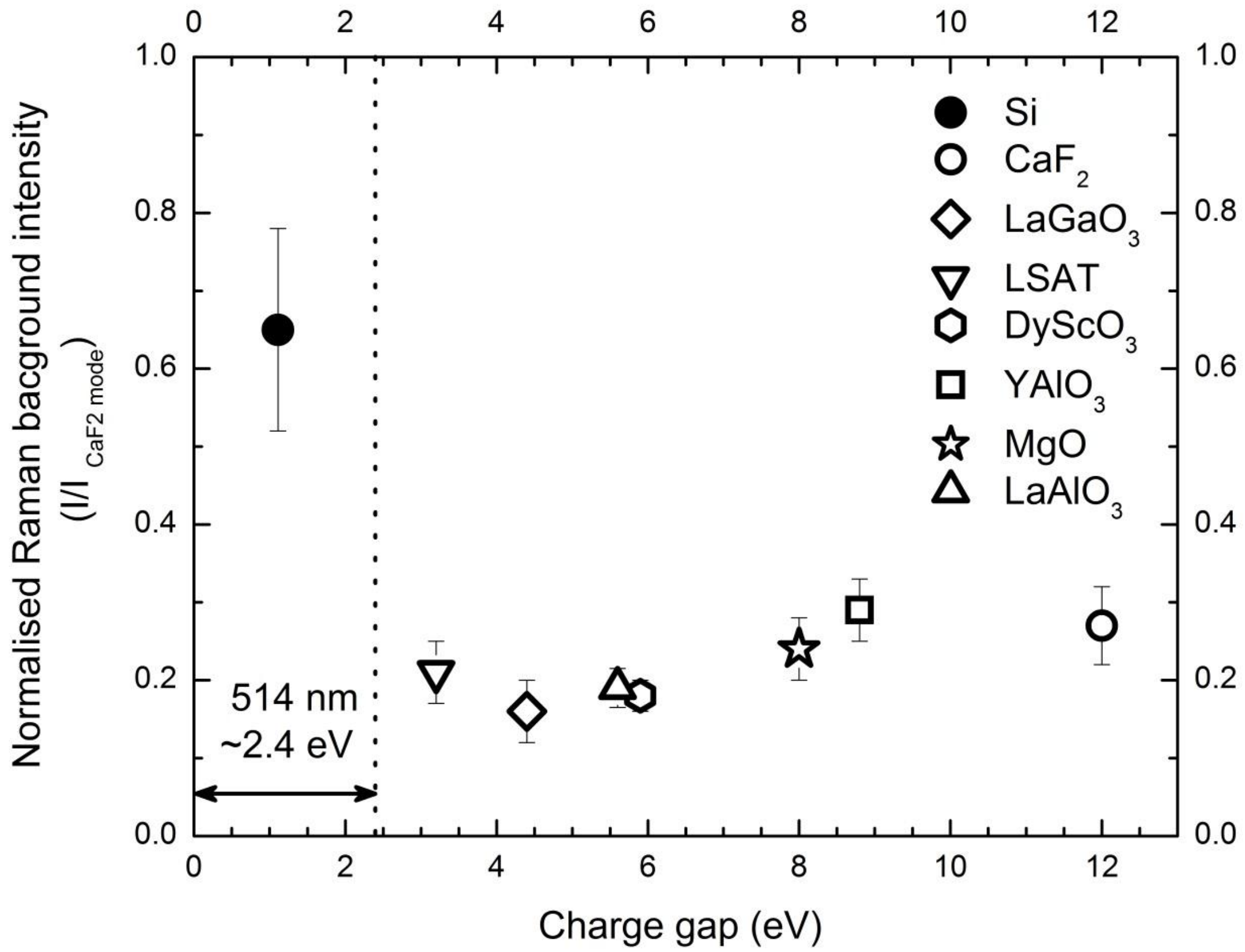

Fig.3 Normalized to the 320cm$^{-1}$-$CaF_2$ mode Raman background vs. substrate's charge gap. The photon energy of the incident laser radiation is indicated by a dashed line. Note that the normalized Raman background intensity is practically constant for the gap values above the photon's energy of the incident laser radiation.